\documentclass[aps,prl,preprint]{revtex4}

\usepackage{graphics}

\makeatletter
\long\def\@makecaption#1#2{%
  \par
  \vskip\abovecaptionskip
  \begingroup
   \small\rmfamily
   \sbox\@tempboxa{%
    \let\\\heading@cr
    #1. #2%
   }%
   \@ifdim{\wd\@tempboxa >\hsize}{%
    \begingroup
     \samepage
     \flushing
     \let\footnote\@footnotemark@gobble
     #1. #2\par
    \endgroup
   }{%
     \global \@minipagefalse
     \hb@xt@\hsize{\hfil\unhbox\@tempboxa\hfil}%
   }%
  \endgroup
  \vskip\belowcaptionskip
}%
\makeatother

\begin{document}

\title{Possible Field-Temperature Phase Diagrams of Two-Band Superconductors with Paramagnetic Pair-Breaking}

\author{Kyosuke Adachi and Ryusuke Ikeda}

\affiliation{Department of Physics, Kyoto University, Kyoto 606-8502, Japan}

\date{\today}

\begin{abstract}
Possible field-temperature superconducting (SC) phase diagrams in two-band quasi-two-dimensional materials with a strong paramagnetic pair-breaking (PPB) are considered theoretically. Attention is paid to the case under a magnetic field {\it perpendicular} to the SC layers and to essential differences from the counterpart in the ordinary single-band material. It is found by examining the $H_{c2}(T)$ curve and the vortex lattices to be realized close to $H_{c2}$ that a PPB-induced SC phase with a spatial modulation parallel to the field tends to occur more easily than in the single-band case, and that a crisscrossing vortex lattice proposed previously can occur in place of the conventional Fulde-Ferrell-Larkin-Ovchinnikov state within a parameter range. The relevance of the obtained results to FeSe and other iron-based superconductors is discussed. 
\end{abstract}

\maketitle

\section*{1. Introduction}

Over the last decade, superconducting (SC) materials showing remarkably strong paramagnetic pair-breaking (PPB) effects have been discovered, and studying the PPB effect in superconductors with unconventional electronic properties is now one of the important theoretical subjects in condensed matter physics. In the quasi-two-dimensional (2D) heavy-fermion superconductors 
CeCoIn$_5$ \cite{Bianchi1} and NpPd$_2$Al$_5$ \cite{Haga}, the SC transition at high fields and hence at low temperatures has been shown to be nearly discontinuous, and, on the basis of this and the temperature dependence of the resulting $H_{c2}(T)$ curve, an extremely strong PPB effect in these materials has been clarified. Furthermore, the strange high-field phase \cite{Tokiwa,Kenzel14} found in CeCoIn$_5$ in a field parallel to the SC layers has been interpreted, by an approach based on a {\it single-band} electronic model, as a vortex lattice corresponding to the appearance of a Fulde-Ferrell-Larkin-Ovchinnikov (FFLO) state \cite{FF,LO} with a PPB-induced spatial modulation parallel to the magnetic field \cite{RI2,HI15}. The appearance of the PPB-induced spatial modulation parallel to the field rather than the corresponding perpendicular modulation \cite{Klein} is a rare event \cite{AI03} in the weak-coupling BCS model and has been interpreted \cite{RI1} as a reflection of the fact that this material at ambient pressure is close to an antiferromagnetic quantum critical point. Although the presence of a FFLO vortex state has also been suggested in CeCoIn$_5$ in a field perpendicular to the $ab$-plane, i.e., the SC layers (${\bf H} \parallel c$) \cite{Kumagai06}, one expects the FFLO state to usually appear in a field parallel to the SC layers in various quasi-2D materials including organics \cite{organics}. Several recent experiments on iron-based superconductors have also suggested SC properties induced by PPB in the fields parallel \cite{KFeAs,LiFeAs,KFeSeEKFAs} and perpendicular \cite{KFeSeEKFAs} to the SC layers such as a suppressed $H_{c2}(T)$ curve at low temperatures. However, the iron-based superconductors are typically {\it two-band} materials, and thus qualitatively different properties based on PPB may be expected to occur in these systems. In fact, clear evidence on the discontinuous $H_{c2}$ transition has not been reported in iron-based materials, except for a few examples \cite{Zocco}. Besides this, the recent thermal conductivity measurement on the superconductor FeSe has led to the detection of a high-field SC phase in ${\bf H} \parallel c$ \cite{Kasahara}. From the large Maki parameter estimated on one of two bands, it is reasonable to expect this high field SC phase to stem from a large PPB effect. 

In this work, we study possible field-temperature phase diagrams of two-band quasi-two-dimensional superconductors in a magnetic field {\it perpendicular} to the layers by constructing the Ginzburg-Landau (GL) free-energy functional based on a microscopic model. In contrast to the single-band case \cite{AI03} where the PPB strength is uniquely characterized by a single Maki parameter $\alpha_{\rm M}$, one cannot predict the resulting phase diagram only on the basis of the PPB strength in multi-band cases where a different $\alpha_{\rm M}$ is defined on a different band. By focusing on a two-band model in which the two $\alpha_{\rm M}$s remarkably differ from each other, we have a couple of categories of typical phase diagrams depending on the details or the breakdown of the parameters determining $\alpha_{\rm M}$s. When the interband coupling is substantial, we find two typical features that do not appear in the single-band case mentioned above. Firstly, we have a wide parameter range in which a PPB-induced spatial modulation characterizing a FFLO state is formed in the direction parallel to the field. In this state, the PPB-induced spatial modulation of a SC order parameter has the following form: $\Delta ({\bf r}) \propto \exp ({i} k z)$ for a FF state and $\Delta ({\bf r}) \propto \cos (k z)$ for a LO state, where $k$ is a modulation wave number and the $z$-axis is parallel to the field. Secondly, over some range of material parameters, we find the situations in which the low-temperature $H_{c2}$ transition is not a discontinuous but a continuous one to a vortex lattice with a modulation parallel to the field. This finding seems to be consistent with the fact that, in the iron-based superconductors, the first-order-like $H_{c2}$ transition has been rarely seen so far even in the cases indicative of a strong PPB effect. On the basis of a previous argument \cite{RIPRLCom} given in relation to phenomena in CeCoIn$_5$, these two features satisfy the condition for the emergence of a crisscrossing vortex lattice \cite{Agter} as the PPB-induced modulated high-field SC state, where, in contrast to a FFLO state, the order parameter has only spatial line nodes. In addition, stable half-flux quanta can emerge in this state. Conversely, if the interband coupling is sufficiently weak, the resulting phase diagram includes, as in the single-band case, a vortex state with a modulation in the plane perpendicular to the applied field at least within the weak-coupling BCS model \cite{Klein,AI03}. 

The outline of this manuscript is as follows. In Sect. 2, the model and methods of theoretical calculations are shown. In Sect. 3, the obtained phase diagrams are presented. In Sect. 4, our results are discussed in relation to the experimental observations. In Appendix, the details of derivations of theoretical expressions are explained. 
 
\section*{2. Methods}

We start with the following quasi-2D two-band Hamiltonian $\mathcal{H}$ in an external field $H$ parallel to the $c$-axis: 

\begin{eqnarray}
{\cal H} &=& {\cal H}_{2 {\rm D}} + {\cal H}_{\rm hop} + {\cal H}_{\rm int} \nonumber \\
&=& \sum_{j=1}^{N_{\rm layer}} \int_\Omega {d}^2{\bf r} \sum_{n=1,2} \sum_{\sigma=\pm1} [\psi_{n}^{\sigma}({\bf r}, j)]^\dagger \{ \varepsilon_n [-{ i} \nabla - e {\bf A}({\bf r}) ] - \sigma I_{n} \} \psi_{n}^\sigma({\bf r}, j) \nonumber \\
&+& \sum_{j=1}^{N_{\rm layer}} \int_\Omega {d}^2{\bf r} \sum_{n=1,2} \sum_{\sigma=\pm1} \frac{J}{2} \Bigg\{ [\psi_{n}^{\sigma} ({\bf r}, j+1) ]^\dagger \psi_{n}^{\sigma}({\bf r}, j) + [\psi_{n}^{\sigma}({\bf r}, j)]^\dagger \psi_{n}^{\sigma}({\bf r}, j+1) \Bigg\} \nonumber \\
&-& \frac{1}{2} \sum_{j=1}^{N_{\rm layer}} \int_\Omega {d}^2{\bf r} \sum_{n,n'} ({U})_{nn'} \sum _{\sigma=\pm1} [\psi_{n}^{\sigma}({\bf r}, j)]^\dagger [\psi_{n}^{-\sigma}({\bf r}, j)]^\dagger \psi_{n'}^{-\sigma}({\bf r}, j)\psi_{n'}^{\sigma}({\bf r}, j). 
\label{startH}
\end{eqnarray}
The first line ${\cal H}_{2{\rm D}}$ and the second line ${\cal H}_{\rm hop}$ in Eq. (\ref{startH}) describe the in-plane and inter-plane motions of quasiparticles in a magnetic field, respectively, while the third term ${\cal H}_{\rm int}$ denotes mutual interactions between quasiparticles. Furthermore, we use the following symbols: $\Omega$ the area of each layer; $N_{\rm layer}$ the number of layers; $\varepsilon_n({\bf p})$ ($n=1$ and $2$) the energy dispersion of a quasiparticle in the $ab$-plane on the $n$-th band; $I_n = g_n \mu_{\rm B} H/{2}$ the Zeeman energy on the $n$-th band; $\sigma$ the spin projection index; $g_n$ the effective $g$-factor on each band; $\mu_{\rm B}$ the Bohr magneton; $J$ the interlayer hopping energy; and $({U})_{nn}=u_n>0$ and $({U})_{12}=({U})_{21}=u_3$ the intraband and interband interaction constants. Note that the $z$-axis is parallel to the $c$-axis, and that ${\bf r}$ and $j$ are the in-plane coordinate and layer number, respectively. 
Concerning each energy dispersion $\varepsilon_n({\bf p})$, cylindrical symmetry is assumed for simplicity. 
The field operator of quasi-particles is defined as $\psi_{n}^\sigma({\bf r}, j) = (\Omega N_{\rm layer})^{-1/2} \sum_{{\bf p},k} \exp[{i}({\bf p}\cdot{\bf r} +kj)] a_{n}^\sigma({\bf p}, k)$ ($-\pi < k \leq \pi$) with the corresponding annihilation operator $a_{n}^\sigma({\bf p}, k)$. Hereafter, the type II limit is assumed so that the magnetic field in the system is approximately equal to the applied field. At least near the resulting $H_{c2}(T)$ line, which is the so-called upper critical field in the mean field approximation, this treatment is safely valid. In addition, except in the case with an extremely large $H_{c2}(0)$, we can use the quasi-classical approximation in which the effect of the magnetic field on the one-particle Green's function on the $n$-th band is represented in terms of a phase factor as 
\begin{equation}
{\cal G}_n^\sigma ({\bf r}, {\bf r}', j-j', \omega_l) 
\simeq 
\exp{\Bigg[{i} e \int_{{\bf r}'}^{\bf r} {d}{\bf l}\cdot{\bf A}({\bf l})\Bigg]} 
\frac{1}{\Omega N_{\rm layer}} \sum_{{\bf p}, k} 
\exp\{{i}[{\bf p}\cdot({\bf r} - {\bf r}') + k (j-j')])\} 
{\cal G}_n^\sigma ({\bf p}, k, \omega_l), 
\end{equation}
where
$
{\cal G}_n^\sigma ({\bf p}, k, \omega_l)
=
[ {i} \omega_l - \xi_n({\bf p}) + \sigma I_n - J {\rm cos}k ]^{-1}
$, 
$\xi_n({\bf p})= \varepsilon_n({\bf p})-\mu$ with chemical potential $\mu$, and $\omega_l$ is the Fermion Matsubara frequency. Below, by the functional integral method and the Hubbard-Stratonovich (HS) transformation, we derive the GL functional and then examine the $H_{c2}(T)$ line and the nature of the mean field transition in a two-band 
system. The ensuing treatment will be explained hereafter separately depending on the sign of $|{U}| \equiv u_1 u_2 - u_3^2$. 

\subsection{(i) $|{U}| > 0$} 

When $|{U}| > 0$, as shown in Appendix, the auxiliary field $\Delta_n({\bf r}, j)$ corresponding to the SC order parameter field $\sum_{n'} ({U})_{nn'} \langle \psi_{n'}^\downarrow({\bf r},j) \psi_{n'}^\uparrow({\bf r},j) \rangle$ can be straightforwardly introduced to obtain the GL functional 
\begin{eqnarray}
\Omega_{\rm GL}
\left[ \Delta_1, \Delta_2 \right] 
&=& 
\Omega_{(0)}
 + 
\sum_j \int {d}^2{\bf r} \sum_{n,n'} 
{\Delta}^*_n \left( {U}^{-1} \right)_{nn'} {\Delta}_{n'}
\nonumber \\
&-&
\frac{T}{2!}
\left< {S'}^{\, 2} \right>_{(0)}^{\rm c}
- \frac{T}{4!}
\left< {S'}^{\, 4} \right>_{(0)}^{\rm c} 
+ {\cal O} \left( {\Delta_1}^{6}, {\Delta_2}^{6} \right), 
\label{eq:GLfcnl}
\end{eqnarray}
where $\Omega_{(0)}$ is the thermodynamic potential in the normal phase, 
\begin{equation}
S' = 
- \int_0^\beta {d}\tau \sum_j \int {d}^2{\bf r} \sum_{n}
\biggl[ 
\Delta_n {\overline \psi}_{n}^\uparrow(\tau) {\overline \psi}_{n}^{\downarrow}(\tau) 
+ {\Delta}^*_n {\psi}_{n}^\downarrow(\tau) {\psi}_{n}^{\uparrow}(\tau) 
\biggr], 
\end{equation}
and $\langle X \rangle_{(0)}^{\rm c}$ denotes the ensemble average of $X$ with respect to the noninteracting part of the action with only the connected diagrams retained. 
We note that here the Fermion fields $\psi_n^\sigma({\bf r}, j, \tau)$ and ${\overline \psi}_n^\sigma({\bf r}, j,\tau)$ are the Grassmann numbers. 

First, let us derive the quadratic (${\cal O}(|\Delta|^2)$) terms. 
In this case, the quadratic term in $\Omega_{\rm GL}$ takes the form 
\begin{equation}
\Omega_{(2)}
=
\sum_{n,n'} \sum_k \int {d}^2{\bf r} \, {\Delta}^*_{n}({\bf r}, k) 
\left[
\left( {U}^{-1} \right)_{nn'} 
 - 
\delta_{nn'} \widehat{K}_{(2)n}({\widehat{\bf \Pi}}, k) 
\right] 
{\Delta}_{n'}({\bf r}, k), 
\label{eq:Omega2}
\end{equation}
where
\begin{equation}
\widehat{K}_{(2)n}({\widehat{\bf \Pi}}, k)
=
\frac{T}{2 \Omega N_{\rm layer}}
\sum_{{\bf p}, \omega_{l}} \sum_{\sigma=\pm1} \sum_{k'} 
{\cal G}_n^{\sigma} ({\bf p}, k', \omega_{l})
{\cal G}_n^{-\sigma} ({\widehat{\bf \Pi}} - {\bf p}, k - k', -\omega_{l}),
\label{eq:K2}
\end{equation}
${\widehat{\bf \Pi}}=-{i}\nabla-2e {\bf A}({\bf r})$ is the gauge-invariant gradient operator, and the Fourier transformation is defined as
$
\Delta_{n}({\bf r}, j)
=
({N_{\rm layer}})^{-1/2} \sum_{k} \exp({{i} kj}) \Delta_{n}({\bf r}, k).
$
Following the route sketched in Appendix, we introduce the parameter $\rho$ and the cutoff $\rho_{\rm cut}$ instead of the energy cutoff, and then we 
obtain 
\begin{eqnarray}
\Omega_{(2)}
&=& 
\sum_{n,n'} \sum_{N=0}^{\infty} \sum_k \int {d}^2{\bf r} \, 
 [{\Delta}^{(N)}_{n}({\bf r}, k)]^* 
\biggl[
\biggl( {U}^{-1} \biggr)_{nn'}
- 
\delta_{n,n'} D_n {f}_{n}^{(N)}(k;T,H) \biggr] 
{\Delta}_{n'}^{(N)}({\bf r}, k) \nonumber 
\\ 
&=&
\sum_{N=0}^{\infty} \sum_k \int {d}^2{\bf r} 
	\Biggl\{
	\biggl[
	{\overline u}_{2}-D_{1}f_{1}^{(N)}(k;T,H)
	\biggr] 
	 |\Delta_{1}^{(N)}({\bf r}, k)|^{2} + 
	\biggl[ 
	{\overline u}_{1} - D_{2}f_{2}^{(N)}(k;T,H)
	\biggr] \nonumber \\
&\times& 
	|\Delta_{2}^{(N)}({\bf r}, k)|^{2} 
 -
	{\overline u}_{3} \bigg[ [{\Delta}_{1}^{(N)}({\bf r}, k)]^* {\Delta}_{2}^{(N)}({\bf r}, k) + {\rm (c. \, c.)} \bigg]
	\Biggr\},
\label{secondOmega2}
\end{eqnarray}
where 
\begin{eqnarray}
f_{n}^{(N)}(k;T,H)
&=& 
2 \pi T \int_{\rho_{\rm cut}}^{\infty} {d} \rho \, \frac{{\rm cos}(2I_n \rho)}{{\rm sinh}(2 \pi T \rho)} \, {\rm J}_0 \biggl[ 2J \rho \, {\rm sin}\biggl(\frac{k}{2}\biggr) \biggr] \, \exp\biggl({-\frac{{v_n}^2 \rho^2}{4 {\lambda_H}^2}} \biggr) \, {\rm L}_N \biggl(\frac{{v_n}^2\rho^2}{2 {\lambda_H}^2} \biggr),
\end{eqnarray}
$\Delta_n^{(N)}({\bf r}, k)$ represents the $N$-th Landau level component of $\Delta_n({\bf r}, k)$, ${\overline u}_j = u_j/|{U}|$, 
${\rm J}_{0}(x)$ represents the zeroth-order Bessel function, and ${\rm L}_N(x)$ represents the $N$-th-order Laguerre polynomial. 
Furthermore, $D_n$ and $v_n$ are the density of states (DOS) at the Fermi surface and the Fermi velocity on the $n$-th band, respectively, and $\lambda_H = (2|e| H)^{-1/2}$ is the magnetic length. 
By diagonalizing the expression of Eq. (\ref{secondOmega2}) and using the resulting eigenvalues $c_\pm^{(N)}$ ($c_+^{(N)} > c_-^{(N)}$) and the `eigenfunctions' $\Delta_\pm^{(N)} = \sum_n w_{\pm,n}^{(N)} \Delta_n^{(N)}$, $\Omega_{(2)}$ may be expressed as 
\begin{equation}
\Omega_{(2)}
=
\sum_{N=0}^{\infty} \sum_k \int {d}^2{\bf r} \left[
c_{+}^{(N)}(k;T,H)|\Delta_{+}^{(N)}({\bf r}, k)|^{2} + c_{-}^{(N)}(k;T,H)|\Delta_{-}^{(N)}({\bf r}, k)|^{2}
\right].  
\label{plusminus}
\end{equation}
We call $\Delta_\pm^{(N)}$ merely as the `$\pm$ component' of $\Delta^{(N)}$. Therefore, if the mean field transition is of the second order, the equation determining $H_{c2}(T)$ is $c^{(N)}_{-}(k;T,H)=0$, or equivalently, 
\begin{equation}
1-\Lambda_{1}f_{1}^{(N)}(k;T,H)-\Lambda_{2}f_{2}^{(N)}(k;T,H)
+ (\Lambda_1 \Lambda_2 - {\Lambda_3}^2) f_{1}^{(N)}(k;T,H)f_{2}^{(N)}(k;T,H)=0,
\label{transeq}
\end{equation}
where $N$ and $k$ should be chosen so that the magnitude of $H_{c2}$ becomes highest for a given $T$, and $\Lambda_1=D_1u_1$, $\Lambda_2=D_2u_2$, and $\Lambda_3=\sqrt{D_1D_2}u_3$ are the dimensionless coupling constants.

Next, we turn to deriving the quartic (${\cal O}(|\Delta|^4)$) term in $\Omega_{\rm GL}$. Formally, this term is expressed as 
\begin{eqnarray}
\Omega_{(4)} &=& 
\frac{1}{2} \sum_{n} \sum_{k_1, k_2, k_3}
\int {d}^2{\bf r} \, 
\widehat{K}_{(4)n} (\{\widehat{\bf \Pi}_{i}}, {k_i\})
{\Delta}^*_{n}({\bf r}_{1}, k_1)
{\Delta}_{n}({\bf r}_{2}, k_2) \nonumber \\
&\times& 
{\Delta}^*_{n}({\bf r}_{3}, k_3)
{\Delta}_{n}({\bf r}_{4}, k_1+k_3 - k_2)
\Big|_{{\bf r}_{i} \rightarrow {\bf r}},
\label{4th}
\end{eqnarray}
where
\begin{eqnarray}
\widehat{K}_{(4)n} (\{ \widehat{\bf \Pi}_{i}, k_i\})
&=& 
\frac{T}{\Omega N_{\rm layer}} 
\sum_{{\bf p}, \omega_l}
\biggl\langle 
{\cal G}_n^\sigma({\bf p}, k, \omega_l) 
{\cal G}_n^{-\sigma}({\widehat{\bf \Pi}}^*_1- {\bf p}, k_1-k, -\omega_l) \nonumber \\
&\times& 
{\cal G}_n^{-\sigma}({\widehat{\bf \Pi}}_2-{\bf p}, k_2-k, -\omega_l) 
{\cal G}_n^\sigma({\widehat{\bf \Pi}}_3^* - {\widehat{\bf \Pi}}_2 + {\bf p}, k_3-k_2+k, 
\omega_l) 
\biggr\rangle_{k,\sigma}, 
\label{eq:K4}
\end{eqnarray}
and $\langle X \rangle_Y$ denotes the average of $X$ with respect to $Y$. Because we are interested in the region close to the resulting $H_{c2}(T)$, we only have to focus on the `$-$ component' $\Delta_-^{(0)}$, defined through the above-mentioned diagonalization, to examine the sign of the GL-quartic term and thus the nature of the mean field $H_{c2}$ transition. Furthermore, for the moment, we will restrict ourselves to the lowest Landau level (LLL) modes of the order parameter, denoted as $\Delta^{(0)}$ below, because, as will be explained in the next section, we often encounter the situations dominated by the LLL modes in the present two-band cases despite the presence of a large paramagnetic pair-breaking. We note that, in the familiar single-band case, a larger paramagnetic pair-breaking makes the vortex state a state dominated by higher Landau level modes \cite{Klein,AI03}. By using the Landau gauge (${\bf A}= B x \widehat{y}$), the order parameter will be written below in the form 
\begin{eqnarray}
\Delta_n({\bf r}, j)
&=&
\frac{w_{-,n}^{(0)}(k;T,H)}{\sqrt{N_{\rm layer}}}
\left[
e^{{i}k j} \Delta^{(0)}_-({\bf r}, k)
+ 
e^{-{i} k j} \Delta^{(0)}_-({\bf r}, -k)
\right]
\nonumber \\
&=&
\frac{w_{-,n}^{(0)}(k;T,H)}{\sqrt{N_{\rm layer} S_{H}}}
\sum_q
u_q({\bf r})
\left[
e^{{i} k j} \, \phi(q,k) + e^{-{i} k j} \, \phi(q,-k)
\right],
\label{op}
\end{eqnarray}
where 
$S_{H}
=
\sqrt{\pi} \lambda_{H} L_y$, 
$L_y$ is the system size in the $y$-direction, and 
$u_q({\bf r}) 
= 
\exp\left[-{(x+q{\lambda_{H}}^2)^2}/(2{\lambda_{H}}^2) + {i} qy \right]$. Using Eq. (\ref{op}) and following the procedures to be explained in Appendix, the quartic term is rewritten in the form 
\begin{eqnarray}
\Omega_{(4)}\Big|_{{\rm LLL},\pm k} 
&=& 
\frac{1}{\sqrt{2}{N_{\rm layer} S_{H}}} \sum_{n,\, \{q_i\}}
[w_{-,n}^{(0)}(k;T,H)]^4 \, 
\delta_{q_1+q_3,q_2+q_4} \, e^{
{-({q_{13}}^2 + {q_{24}}^2)}{\lambda_H}^2/{4} 
} \Big[ 
V_{n}(\, \{q_i\}, k; T, H) \, 
\phi^{*}(q_1, k) \nonumber \\
&\times& \phi(q_2, k) 
\phi^{*}(q_3, k) 
\phi(q_4, k) 
+ 
\widetilde{V}_{n}(\, \{q_i\}, k; T, H) \, 
\phi^{*}(q_1, k)
\phi(q_2, k)
\phi^*(q_3, -k)
\phi(q_4, -k) \nonumber \\
&+& 
\widetilde{V}'_{n}(\, \{q_i\}, k; T, H) \, 
\phi^*(q_1,k)
\phi(q_2,-k)
\phi^*(q_3,-k)
\phi(q_4,k) + 
(k \leftrightarrow -k)
\Big],
\label{nonlocal}
\end{eqnarray}
where $q_{ij}=q_i-q_j$, 
\begin{eqnarray}
V_{n} &=& 
2 \pi T D_n 
\int_0^{\infty} {d}\rho_1 {d}\rho_2 {d}\rho_3 \, 
\frac{{\rm cos}[2I_n (\rho_1+\rho_2+\rho_3)]}{{\rm sinh}[2 \pi T(\rho_1+\rho_2+\rho_3)]} 
{\rm J}_0\biggl(2 J (\rho_1+\rho_2+\rho_3) \, {\rm sin}\biggl(\frac{k}{2} \biggr) 
\biggr) 
\nonumber \\ 
&\times& \biggl\langle {\rm Re}  
\exp{[B(\{q_i, \beta_{n,i}\})]} |_{\rho_4=0}  
\biggr\rangle_{\widehat {\bf p}},
 \nonumber \\
\widetilde{V}_{n} 
&=& 
2 \pi T D_n
\int_0^{\infty} {d}\rho_1 {d}\rho_2 {d}\rho_3 \, 
\frac{{\rm cos}[2 I_n (\rho_1+\rho_2+\rho_3)]}{{\rm sinh}[2 \pi T(\rho_1+\rho_2 + \rho_3)]} 
{\rm J}_0 
\biggl( 
2 J \sqrt{(\rho_1+\rho_2-\rho_3 \, {\rm cos}(k))^2+(\rho_3 \, {\rm sin}(k))^2} \, \nonumber \\
&\times& {\rm sin}\biggl(\frac{k}{2} \biggr) 
\biggr) \biggl\langle
{\rm Re}  
\exp{[B(\{q_i, \beta_{n,i}\})]}|_{\rho_4=0}  
\biggr\rangle_{\widehat {\bf p}}, 
\end{eqnarray}
$\widetilde{V}'_n$ is $\widetilde{V}_n$ with $\rho_1$ and $\rho_3$ exchanged with each other, and $B(\{q_i,\beta_{n,i}\})$ will be defined in Appendix. If we assume that the $k$- and $q_i$-dependences of both $V_{n}$ and $\widetilde{V}_{n}$ are weak and negligible, the quartic term is expressed in the local form 
\begin{eqnarray}
\Omega_{(4)}^{{\rm local}}\Big|_{{\rm LLL},\pm k}
=
\sum_j \int {d}^2{\bf r}
\left\{
\sum_n
[w_{-,n}^{(0)}(k;T,H)]^4
V_{n}^{\rm local}(T,H)
\right\}
|\Delta_-^{(0)}({\bf r}, j)|^4,
\end{eqnarray}
where
\begin{eqnarray}
V_{n}^{\rm local}(T,H)
&=&
2 \pi T D_n
\int_0^{\infty}{d}\rho_1 {d}\rho_2 {d}\rho_3 
\frac{{\rm cos}[2I_n (\rho_1+\rho_2+\rho_3)]}{{\rm sinh}[2 \pi T(\rho_1+\rho_2+\rho_3)]}
\left<
{\rm Re}
\left[
e^{B(\{0\},\{\beta_{n,i}\})}\Big|_{\rho_4=0}
\right]
\right>_{\widehat {\bf p}}.
\label{eq:Vlocal}
\end{eqnarray}

By using $\Omega_{(4)}^{\rm local}|_{{\rm LLL},\pm k}$, we estimate the region in which a first-order transition occurs; if the quartic term is positive on the line where the quadratic term vanishes, a second-order transition occurs; otherwise, a first-order transition is expected to occur at a higher field.

\subsection{(ii) $|{U}|<0$}

When $|{U}|<0$, either of the two eigenvalues of ${U}$ is negative, and then the interaction Hamiltonian takes the form 
\begin{equation}
-\sum_{n,n'} u_{nn'} \, [\psi_{n}^{\uparrow} ({\bf r}, j)]^\dagger [\psi_{n}^{\downarrow}({\bf r}, j)]^\dagger \psi_{n'}^{\downarrow}({\bf r}, j) \psi_{n'}^{\uparrow}({\bf r}, j)
= 
-u_+ \Phi_+^\dagger \Phi_+ + |u_-| \Phi_-^\dagger \Phi_-, 
\label{linear}
\end{equation}
where $\Phi_\pm$ are appropriate linear combinations of $\psi_n^\downarrow \psi _n^\uparrow$. 
Because $u_-$ is negative, the second term of Eq. (\ref{linear}) may be regarded as a contribution implying a repulsive interaction, and thus we neglect the term from now on. 
Then, as in the case of $|{U}|>0$, performing the HS transformation leads to a GL functional. Then, the corresponding quadratic term $\Omega_{(2)}$ [corresponding to Eq. (\ref{plusminus})] becomes
\begin{equation}
\Omega_{(2)}
=
\sum_{N=0}^{\infty} \sum_k 
\left[
\frac{1}{u_+}
- 
\sum_n 
{\widetilde{w}_n}{^2}
D_n 
f_n^{(N)}(k;T,H)
\right]
\int {d}^2{\bf r} \, |\widetilde{\Delta}_{+}^{(N)}({\bf r}, k)|^{2},
\end{equation}
where $\widetilde{\bf w}$ is the eigenvector that belongs to the eigenvalue $u_+$, and $\widetilde{\Delta}_{+}^{(N)} = \sum_n \widetilde{w}_{n} \Delta_n^{(N)}$. We note that $\widetilde{\Delta}_{+}^{(N)}$ is different from $\Delta_+^{(N)}$ defined in the $|{U}|>0$ case. 
The quartic term $\Omega_{(4)}$ has the same form as in the $|{U}|>0$ case [see Eqs. (\ref{nonlocal}) and (\ref{eq:Vlocal})], if $w_{-,n}^{(0)}(k;T,H)$ and $\Delta_-^{(0)}$ in $|{U}| > 0$ case are replaced by $\widetilde{w}_n$ and $\widetilde{\Delta}_{+}^{(0)}$, respectively.

\section*{3. Results}
\label{sec:result}

By numerically calculating the coefficients of the quadratic and quartic GL terms, we estimate the $H_{c2}(T)$ line in the case where the PPB effect is strong on one of the two bands. Below, we will focus especially on this case by taking account of the SC properties seen in FeSe (see Sect. 1) \cite{Kasahara}. 
We represent the relative strength of the paramagnetic effect to the orbital effect on each band via the parameter $\alpha_n=Cg_n T_{c0}/(m_0{v_n}^2) \propto g_n/{v_n}^2$ proportional to the Maki parameter $\alpha_{\rm M}$, where $C=\pi/2\exp(\gamma) \simeq 0.88$, $\gamma$ is the Euler constant, $m_0$ is the electron mass, and $T_{c0}$ is the mean field transition temperature in the absence of the magnetic field. 

Let us start with reviewing the single-band limit in which $\Lambda_2=\Lambda_3=0$, and the PPB on the only attractive band is given by $\alpha_1=g_1\mu_{\rm B} H_1^{({\rm orb})}/(2 \pi T_{c0})$, where $H_1^{({\rm orb})}$ is $H_{c2}(0)$ in the single-band case with no PPB effect included. In the weak PPB limit where $\alpha_1 \ll 1$, the orbital pair-breaking is dominant, and the second-order $H_{c2}$ transition to a conventional vortex lattice with no additional spatial modulation described in LLL occurs. Conversely, when the PPB is sufficiently strong with, say, $\alpha_1 \geq 1.4$, the vortex lattice just below the $H_{c2}(T)$ line is dominated by higher LLs at low but finite temperatures at least within the weak-coupling BCS model although the $H_{c2}$ transition is of the second order. This higher LL vortex state includes additional spatial modulations accompanied by antivortices in the plane perpendicular to the field. Depending on material parameters such as disorder strength, this higher LL transition is replaced by a {\it discontinuous} $H_{c2}$ transition to the ordinary vortex lattice described in LLL, which may have a PPB-induced spatial modulation only in the direction parallel to the field. 

Hereafter, we will focus on the case in which PPB is strong on band 1 and weak on band 2 by keeping the relation $\alpha_2=\alpha_1/16$, and, regarding the intraband couplings, the values $\Lambda_1=0.2$ and $\Lambda_2=0.15$ will be assumed. The interlayer hopping $J$ will be fixed to $10 \pi T_{c0}$ in all of our computations. Furthermore, to reduce the number of material parameters appearing in our theoretical expressions, the in-plane Fermi momenta of the two bands are assumed to be the same so that $D_2/D_1=m_2^*/m_1^*$, where $m_n^*$ is the effective mass on each band. Qualitatively, theoretical results are unaffected by this simplification. Then, we have the relation $\alpha _1/\alpha_2 ={g_1 {v_2}^2}/(g_2 {v_1}^2)={g_1 {m_1^*}^2}/(g_2 {m_2^*}^2)$. Note that the ratio $D_1/D_2$ can be changed while $\alpha_1/\alpha_2$ is fixed. Hence, even if $\alpha_1$ and $\alpha_2$ are fixed, the character of the $H_{c2}$ transition may be changed for different values of the DOSs $D_1$ and $D_2$. 

Under such a parametrization, we have three control parameters that may affect the resulting phase diagram, i.e., the interband coupling $|\Lambda_3|$, the band asymmetry $m_1^*/m_2^*$ in the effective mass (or the density of states), and another band asymmetry $g_1/g_2$ in the Zeeman term. Roughly speaking, $g_1/g_2$ determines the relative strength of only PPB between the two bands, while $m_1^*/m_2^*$ also affects the ratio of the coherence lengths, i.e., the orbital pair-breaking. 

First, we examine the case with a vanishingly weak interband coupling so that $|\Lambda_3|\ll \Lambda_1$ and $\Lambda_2$. Clearly, the SC transition in this limiting case is dominated by the states on one band. If the intraband couplings $\Lambda_1$ and $\Lambda_2$ are almost equal to each other, the band with a smaller PPB effect dominates in high fields. In the present case where the difference $|\Lambda_1 - \Lambda_2|/\Lambda_1$ is of order unity, however, the situation is close to the above-mentioned single-band limit with $\Lambda_2=\Lambda_3=0$, and the more attractive electronic states on band 1 largely determine the superconductivity despite the relation $\alpha_1 \gg \alpha_2$. In any case, the behaviors in the high-field and low-temperature region are similar to those in the single-band case \cite{com1}. First, the transition character depends only on the value of $\alpha_1$ and does not depend on its breakdown. Second, a first-order transition within the LLL or the transition to a vortex lattice dominated by a higher Landau level occurs at such low temperatures in the $H$-$T$ phase diagram \cite{AI03}. 

%%%%%%%%%%%%%%%%%%%%%% 
\begin{figure}[t]
\scalebox{0.6}[0.6]{\includegraphics{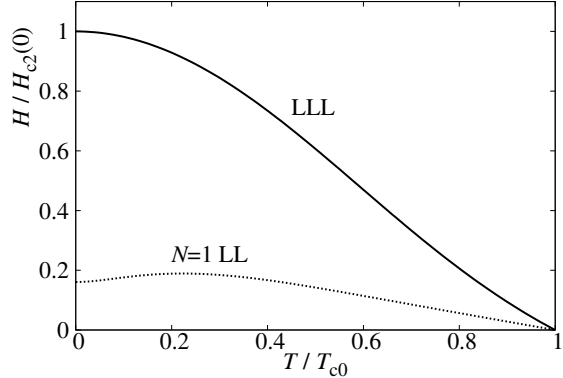}}
%{}
\caption{$H_{c2}(T)$ (black solid) curve for $m_1^*=4m_2^*$, $g_1=g_2$, and $\alpha_1=16 \alpha_2=0.5$. The coupling constants $\Lambda_1=0.2$, $\Lambda_2=0.15$, and $\Lambda_3=0.15$ are used. The nature of the transition on $H_{c2}(T)$ is of the second order at any field, and, because of the large asymmetry in the Fermi velocities on the two bands, the $H_{c2}$ line has a positive curvature in intermediate fields. The $N=1$ LL (black dotted) curve at which $c_-^{(1)}$ changes its sign is drawn for comparison and is not a transition line. }
\label{fig:weak}
\end{figure}
%%%%%%%%%%%%%%%%%%%%%% 

In contrast, in the case with a moderately large interband coupling where $|\Lambda_3| \sim \Lambda_1$ or $|\Lambda_3| \sim \Lambda_2$, a two-band effect appears, and the character of the $H_{c2}$ transition depends not only on the value of $\alpha_n$ but also on their breakdown. First, we will explain the case in which the band asymmetry is due to that of the effective mass for the fixed interband coupling $\Lambda_3=0.15$. Note that this case belongs to the category with $|{U}| > 0$. In the limiting case with no PPB, the $H_{c2}(T)$ curve in this case is well known and has an inflection in intermediate fields \cite{Gurevich}. Such an inflection of $H_{c2}(T)$ is also seen in Fig. \ref{fig:weak}, where the weak PPB has been assumed by setting $\alpha_1=0.5$. In this and the ensuing figures, the temperature is measured in the unit of $T_{c0}$, while the magnetic field is measured in the unit of $H_{c2}(0)$. 
In each figure, the black-solid line denotes the second-order transition line in the mean field approximation on which nucleation of the order parameter in LLL occurs, and the black-dotted curve is the corresponding one in the next lowest ($N=1$) Landau level (LL) obtained by neglecting the presence of the SC ordering in LLL. In the present situation where the LLL ordering occurs at a higher field, no real transition occurs on this line associated with the $N=1$ LL modes. 

%%%%%%%%%%%%%%%%%%%%%% 
\begin{figure}[t]
\scalebox{0.6}[0.6]{\includegraphics{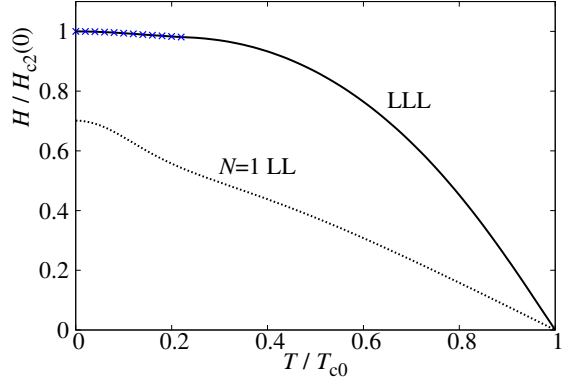}}
%{}
\caption{(Color online) The $H_{c2}(T)$ curve (black solid line with blue cross points) following from larger $\alpha_n$ values, $\alpha_1=16\alpha_2=3.0$. Values of other parameters are the same as those in Fig. \ref{fig:weak}, and thus the band asymmetry in $\alpha_n$ arises from that of the effective mass. The $H_{c2}(T)$ curve is defined in LLL. On its black line portion, $c_-^{(0)}(0)$ changes its sign, and a second-order $H_{c2}$ transition in the mean field approximation occurs between the conventional vortex lattice and the normal phase. On the other hand, the $H_{c2}$ transition occurs as a first-order one in the region specified by blue cross points. For reference, the $N=1$ LL (black dotted) curve is also shown, as in Fig. \ref{fig:weak}. } 
\label{fig:p}
\end{figure}
%%%%%%%%%%%%%%%%%%%%%

Next, we show an example with larger PPB strengths in Fig. \ref{fig:p}. As in the single-band case, the $H_{c2}(T)$ curve takes a suppressed form. However, higher LL vortex states with a PPB-induced spatial modulation perpendicular to the field are not realized even in high fields in contrast to the single-band case \cite{Klein,AI03}. 
Both the $H_{c2}$ line and high-field vortex state are described in LLL. At the temperatures indicated by the blue cross points, the $H_{c2}$ transition is of the first order because the coefficient of the $|\Delta|^4$ term is negative there. Furthermore, the high-field vortex state might be accompanied by a spatial modulation {\it parallel} to the field. However, in Fig. \ref{fig:p}, the region of a modulated phase has not been identified. Further, we have verified that such a phase diagram with a first-order $H_{c2}$ transition defined within LLL is also obtained when $|\Lambda_3|$ is larger so that $|{U}| < 0$. It does not appear that the sign change of $|{U}|$ leads to a qualitative change in the resulting phase diagram. 
Although the $H_{c2}$ transition under a similar situation where the interband coupling is dominant ($|\Lambda_3| \gg \Lambda_1, \Lambda_2$) and the band asymmetry $v_2 / v_1$ is large has been discussed elsewhere \cite{Gurevich}, the possibility of the first-order $H_{c2}$ transition has not been examined there. 

%%%%%%%%%%%%%%%%%%%%%% 
\begin{figure}[t]
\scalebox{0.6}[0.6]{\includegraphics{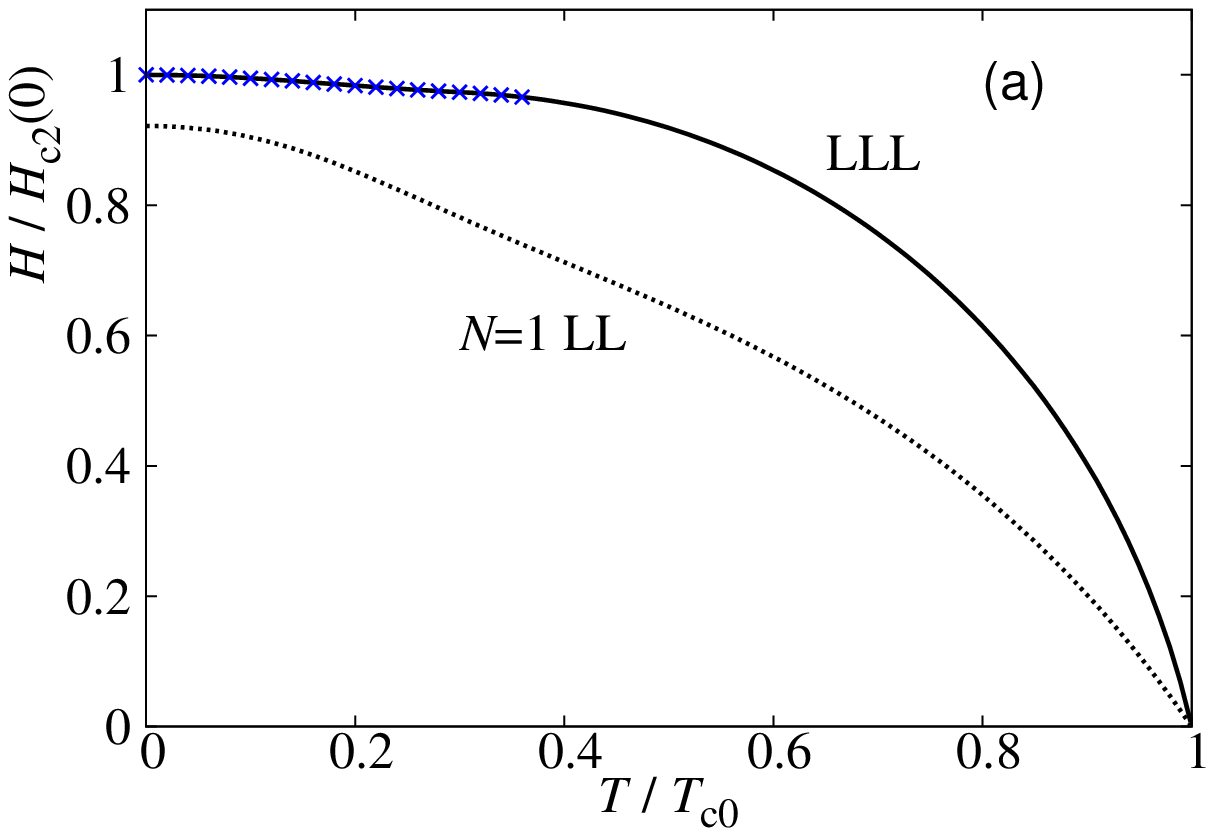}}
\scalebox{0.6}[0.6]{\includegraphics{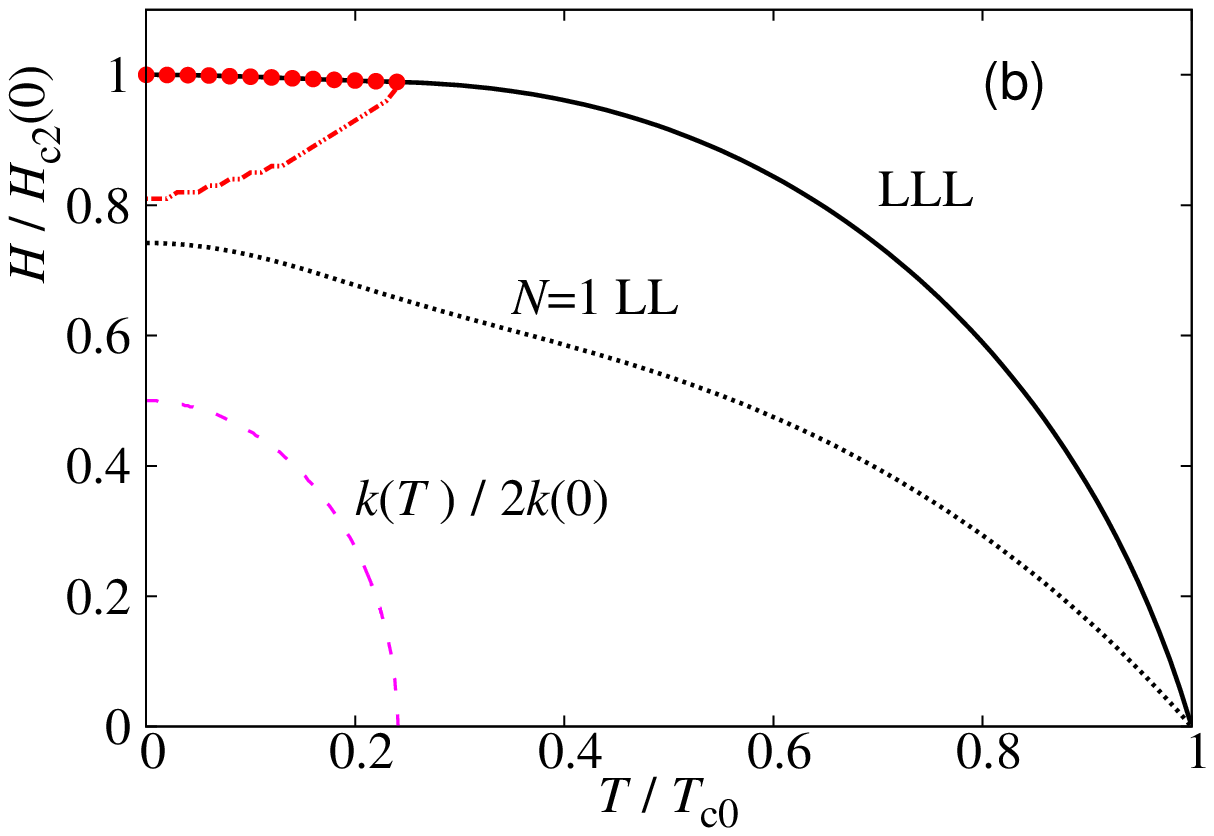}}
\scalebox{0.6}[0.6]{\includegraphics{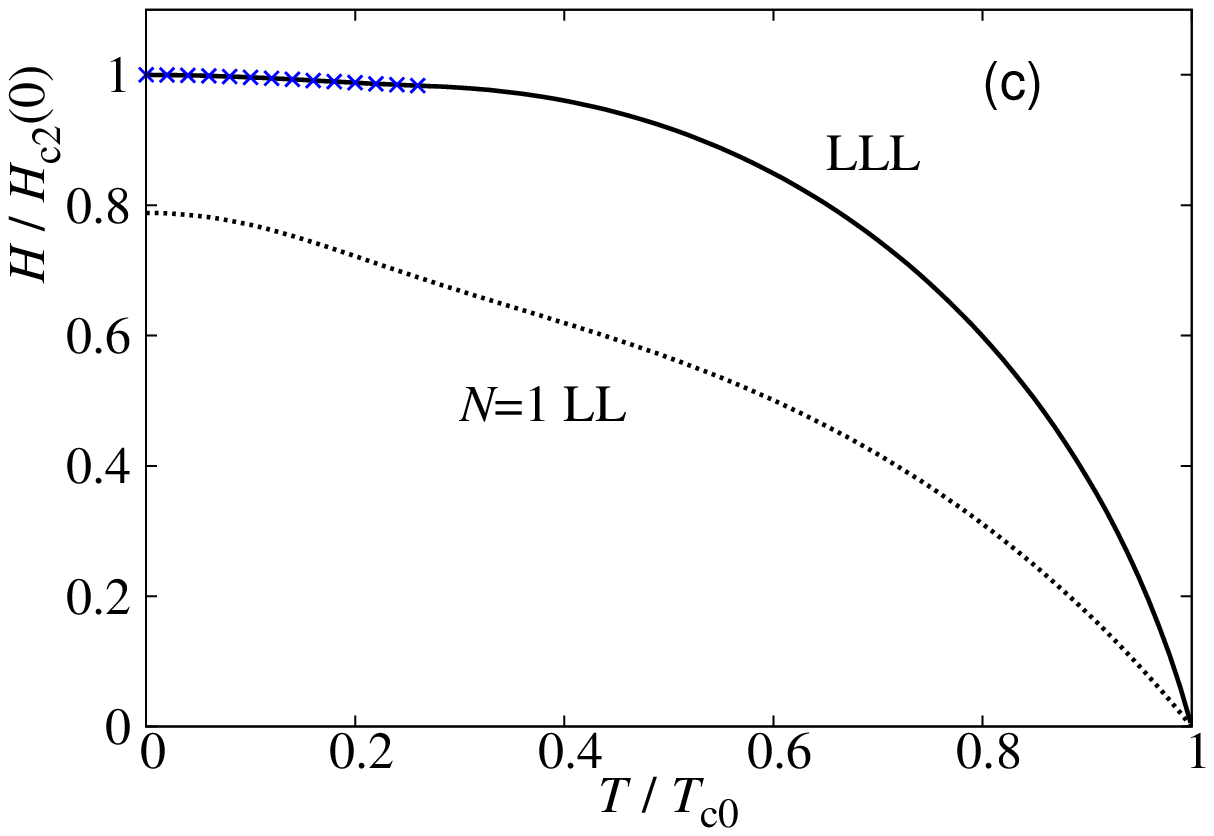}}
%{}
\caption{(Color online) $H_{c2}(T)$ curves (black solid line with red circles or blue cross points) following from the same $\alpha_n$ values as in Fig. \ref{fig:p}. Differing from Fig. \ref{fig:p}, the band asymmetry is assumed here to arise primarily from the Zeeman energy: the parameters describing the breakdown of $\alpha_n$s are chosen as $m_1^*=2m_2^*$ and $g_1=4g_2$ in this case. The intraband couplings are the same as those in Fig. \ref{fig:p}, while only $\Lambda_3$ is changed as (a) 0.05, (b) $0.15$, and (c) $0.4$. For comparison, the $N=1$ LL curve is also shown in each case as in Fig. \ref{fig:weak}, and the transition curve, estimated roughly with no calculation, between the modulated high-field phase and the ordinary vortex lattice is drawn with the red dashed curve. In all cases, the $H_{c2}(T)$ line is defined in LLL, and the nature of the transition is continuous on the black solid curves. On the blue cross points in (a) and (c), the transition is discontinuous, while, on the red circles in (b), the $H_{c2}$ transition becomes the second order to a vortex lattice with a modulation parallel to the field and $c$-axis. The pink-dashed curve expresses the temperature dependence of the magnitude of the modulation wavevector of the high-field SC phase on $H_{c2}(T)$. }
\label{fig:ex}
\end{figure}
%%%%%%%%%%%%%%%%%%%%%

Now, we will explain the case in which the band asymmetry is due primarily to that of the Zeeman energy. The dependences of the phase diagrams belonging to this case on the interband coupling $|\Lambda_3|$ are shown in Fig. \ref{fig:ex}. The figure (a) is a result for a relatively small $\Lambda_3$ and, reflecting the result in the single-band case, the instability line to the $N=1$ LL vortex state is close to the $H_{c2}(T)$ curve defined in LLL. As in Fig. \ref{fig:p}, the $H_{c2}$ transition is discontinuous in the low-temperature region indicated by blue cross points. With increasing interband coupling, the tendency that the LLL modes of the order parameter describe the vortex lattice in high fields is enhanced. In the region where $|\Lambda_3|$ is on the order of $\Lambda_2$, however, the figure (b) shows that the $H_{c2}$ transition remains to be the second order even in a low-temperature limit [see the red circles in (b)]. This low-temperature second-order transition is different from the ordinary one occurring between the familiar vortex lattice and the normal phase in that the ordered phase below $H_{c2}$ is a vortex lattice with a PPB-induced modulation parallel to the field ${\bf H} \parallel c$. The fact that, in the present quasi-two-dimensional system with the simple cylindrical Fermi surfaces, a spatial modulation parallel to the $c$-axis tends to be created is surprising from the viewpoint based on the single-band model and should be regarded as one of the consequences of the multi-band electronic structure. Details of a possible modulated state in this case will be discussed in Sect. 4. 

On the other hand, as $|\Lambda_3|$ becomes significantly large, the $H_{c2}$ transition again becomes discontinuous (mean-field first-order) in the low-temperature region indicated by the blue cross points, and no higher LL vortex lattice tends to appear even at lower temperatures. Furthermore, note that Fig. \ref{fig:ex}(c) belongs to the category with $|{U}| < 0$. Although we have not examined the details of the SC order parameter just below the discontinuous transition line $H_{c2}(T)$, a spatial modulation parallel to the $c$-axis of the SC order parameter similar to that of the LO state is expected to appear there. 
Note that the case with a sufficiently large interband coupling is qualitatively different from the single-band case.

\section*{4. Summary and Discussion}

In this paper, we have attempted to understand the high-field superconductivity in quasi-two-dimensional two-band materials with a remarkable paramagnetic pair-breaking. 
Usually, the paramagnetic pair-breaking effect is expected to occur, even in two-band materials, in a situation with negligibly weak orbital pair-breaking \cite{Mizushima} such as in the field configuration parallel to the conducting layers. In the present work, we have focused rather on the ${\bf H} \parallel c$ case and have found new aspects on two-band superconductivity in high magnetic fields, which cannot be seen in the single-band superconductivity. In the single-band limit, the shape of the $H_{c2} (T)$ curve and the character of the $H_{c2}$ transition depend only on the parameter $\alpha \propto g/|{\bf v}|^2$ proportional to the Maki parameter, which corresponds to the paramagnetic effect relative to the orbital effect, where $g$ is the effective $g$-factor and ${\bf v}$ is the Fermi velocity. 
Thus, the mechanism of the pair-breaking depends only on the ratio of the paramagnetic effect relative to the orbital one. 
In two-band systems, however, a nonvanishing interband coupling $\Lambda_3$ induces new dependences of the $H_{c2}$ transition on material parameters. Not only the $\alpha$-value on each band but also dependences on the parameters describing asymmetries between the two bands can change the nature of the high-field superconductivity. In other words, PPB can be regarded as acting independently of the orbital pair-breaking in two-band superconductors. For instance, a strong PPB is not equivalent to a weak orbital pair-breaking in these systems. 

To elucidate the parameter dependences of the phase diagram, we have changed the band asymmetries in the Fermi velocity (or the effective mass) and Zeeman energy (or the $g$-factor) separately and have investigated their dependences on the interband coupling $|\Lambda_3|$. The former band asymmetry primarily affects the orbital pair-breaking, while the latter measures a difference in PPB between the two bands. It is found that, when the band asymmetry in the Fermi velocity is dominant, the $H_{c2}$ transition at low temperatures tends to become a first order one, and that an additional PPB-induced spatial modulation in the vortex state below the first-order $H_{c2}$ transition can occur only in the direction parallel to the field and hence to the $c$-axis. On the other hand, when the band asymmetry and thus the difference between the Maki parameters defined on the two bands stems mainly from that in the Zeeman energy, we find that, for moderately large $|\Lambda_3|$ values, the $H_{c2}$ transition at low temperatures remains to be continuous, while the PPB-induced spatial modulation just below $H_{c2}$ at low temperatures can occur only along the field $\parallel c$. As pointed out previously \cite{RIPRLCom}, these features correspond to the necessary conditions for the realization of the crisscrossing vortex lattice proposed in Ref. 20. Therefore, a new high field SC phase can be expected to occur in quasi-two-dimensional two-band materials in ${\bf H} \parallel c$. On the other hand, when the $H_{c2}$ transition at low temperatures is of first order, the ordinary Larkin-Ovchinnikov state \cite{LO} modulating along the $c$-axis may occur just below $H_{c2}(T)$. The present results might be relevant to the recent observation of an unexpected high field SC phase in FeSe \cite{Kasahara}. To identify the observed high field phase in FeSe and, if any, the corresponding one in other iron-based superconductors with a theoretical one properly, further model calculations and studies on the fluctuation effect are needed.

\section*{Appendix}

First, we explain how we use the HS transformation and how we obtain the GL functional in the form of Eq. (\ref{eq:GLfcnl}).
Next, the details of the derivation of the GL quadratic term $\Omega_{(2)}$ and the quartic one $\Omega_{(4)}$ are explained. The treatment to derive $\Omega_{(2)}$ and $\Omega_{(4)}$ given below closely follows that in the previous work \cite{AI03}.

\subsection{Hubbard-Stratonovich transformation for interaction term}

We show how we obtain the GL functional in the form of Eq. (\ref{eq:GLfcnl}).
Here, we consider the case of $|{U}|>0$.
By using the functional integral method, from the two-band Hamiltonian (Eq. (\ref{startH})), we can write the partition function for grand-canonical ensemble as
\begin{eqnarray}
\mathcal{Z}
=
\int\mathcal{D}\overline{\psi}
\mathcal{D}\psi\,
e^{-S_{(0)}[\overline{\psi},\psi]-S_{\mathrm{int}}[\overline{\psi},\psi]},
\label{eq:partfcn}
\end{eqnarray}
where $\psi^\sigma_n(\mathbf{x},j,\tau)$ and $\overline{\psi}^\sigma_n(\mathbf{x},j,\tau)$ are the Grassmann numbers that represent fields of quasi-particles and $\tau$ represents imaginary time.
The noninteractive part of the action is given as
\begin{eqnarray}
S_{(0)}
=
\int_0^\beta{d}\tau\sum_j\int{d}^2\mathbf{r}\sum_{n,\sigma}
\overline{\psi}^\sigma_n(\mathbf{r},j,\tau)
\left[
\frac{\partial}{\partial\tau}+\xi_n(-i\nabla-e\mathbf{A}(\mathbf{r}))-\sigma I_n+\widehat{J}
\right]
\psi^\sigma_n(\mathbf{r},j,\tau),
\end{eqnarray}
where for simplicity, we define $\widehat{J}$ as $\overline{\psi}_n^\sigma(j)\widehat{J}\psi_n^\sigma(j)=J[\overline{\psi}_n^\sigma(j+1)\psi_n^\sigma(j)+(j\leftrightarrow j+1)]/2$.
On the other hand, the interactive part is given as
\begin{eqnarray}
S_{\mathrm{int}}
=
-\int_0^\beta{d}\tau\sum_j\int{d}^2\mathbf{r}\sum_{n,n'}
\left({U}\right)_{nn'}
\overline{\psi}_n^\uparrow(\mathbf{r},j,\tau)\overline{\psi}_n^\downarrow(\mathbf{r},j,\tau)
\psi_n^\downarrow(\mathbf{r},j,\tau)\psi_n^\uparrow(\mathbf{r},j,\tau).
\end{eqnarray}

Now, to integrate formally quasi-particles' degrees of freedom, we introduce auxiliary complex fields $\Delta_1(\mathbf{r},j,\tau)$ and $\Delta_2(\mathbf{r},j,\tau)$ corresponding to the SC order parameter fields, and transform the interactive part in Eq. (\ref{eq:partfcn}) as
\begin{eqnarray}
e^{-S_{\mathrm{int}}}
=
\int\mathcal{D}\Delta^*\mathcal{D}\Delta
\exp
\Bigg\{
-\int_0^\beta{d}\tau\sum_j\int{d}^2\mathbf{r}
&&
\Bigg[
\sum_{n,n'}\Delta_n^*\left({U}^{-1}\right)_{nn'}\Delta_{n'}
\nonumber \\
&&-
\sum_n
\left(
\Delta_n\overline{\psi}_n^\uparrow\overline{\psi}_n^\downarrow
+
\Delta_n^*\psi_n^\downarrow\psi_n^\uparrow
\right)
\Bigg]
\Bigg\}.
\end{eqnarray}
Hereafter, because we want to obtain the GL functional that corresponds to the mean field theory, we neglect the quantum fluctuation or $\tau$-dependence of $\Delta_n$.
As a result, we get the partition function as
\begin{eqnarray}
\mathcal{Z}
=
\int\mathcal{D}\Delta^*\mathcal{D}\Delta
\exp
\Bigg[
-\beta\sum_j\int{d}^2\mathbf{r}
\sum_{n,n'}\Delta_n^*\left({U}^{-1}\right)_{nn'}\Delta_{n'}
\Bigg]
\int\mathcal{D}\overline{\psi}\mathcal{D}\psi\,
e^{-S_{(0)}[\overline{\psi},\psi] - S'[\overline{\psi},\psi,\Delta^*,\Delta]},
\label{eq:partfcn2}
\end{eqnarray}
where
$
S'
=
-\int_0^\beta{d}\tau\sum_j\int{d}^2\mathbf{r}\sum_n
\left(
\Delta_n\overline{\psi}_n^\uparrow\overline{\psi}_n^\downarrow
+
\Delta^*_n\psi_n^\downarrow\psi_n^\uparrow
\right).
$

Then, by using the linked-cluster theorem, we can transform Eq. (\ref{eq:partfcn2}) as
\begin{eqnarray}
\mathcal{Z}
=
\int\mathcal{D}\Delta^*\mathcal{D}\Delta
\exp
\Bigg\{
-\beta
\Bigg[
\Omega_{(0)}
+
\sum_j\int{d}^2\mathbf{r}
\sum_{n,n'} & \Delta_n^* & (\mathbf{r},j)\left({U}^{-1}\right)_{nn'}\Delta_{n'}(\mathbf{r},j)
\nonumber \\
&&-
T
\left(
\left<
e^{-S'[\overline{\psi},\psi,\Delta^*,\Delta]}
\right>_{(0)}^{\mathrm{c}}
-
1
\right)
\Bigg]
\Bigg\},
\label{eq:partfcn3}
\end{eqnarray}
where $\Omega_{(0)}=-T\ln[\int\mathcal{D}\overline{\psi}\mathcal{D}\psi\exp(-S_{(0)}[\overline{\psi},\psi])]$ that represents the thermodynamic potential in the normal phase and $\left<X\right>_{(0)}^{\mathrm{c}}$ denotes the ensemble average of $X$ with respect to the non-interactive part of the action with only the connected diagrams retained.
Finally, expanding $\exp(-S')$ with respect to $S'$ in Eq. (\ref{eq:partfcn3}), we obtain the GL functional in the form of Eq. (\ref{eq:GLfcnl}):
\begin{eqnarray}
\Omega_{\mathrm{GL}}
\left[\Delta^*,\Delta\right]
&=&
\Omega_{(0)}
+
\sum _j \int {d}^2\mathbf{r} \sum_{n,n'}
{\Delta}^*_n \left( {U}^{-1} \right)_{nn'} {\Delta}_{n'}
\nonumber \\
&-&
\frac{T}{2!}
\left< {S'}^{\, 2} \right>_{(0)}^{\mathrm{c}}
-\frac{T}{4!}
\left< {S'}^{\, 4} \right>_{(0)}^{\mathrm{c}}
+ \mathcal{O} \left( {\Delta_1}^{6}, {\Delta_2}^{6} \right),
\end{eqnarray}
where the minimum point of $\Omega_{\mathrm{GL}}$ corresponds to an equilibrium state at the mean field level.

\subsection{Calculation of $\Omega_{(2)}$}

First, $\widehat{K}_{(2)n}(\widehat{\bf \Pi}, k)$ defined in Eq. (\ref{eq:K2}) will be rewritten in a convenient form for numerical calculations. After integrating over $\xi_n({\bf p})$ and exponentiating the resulting denominator in terms of the integral 
\begin{equation}
\frac{1}{X} = \int_0^\infty d\rho \exp(-\rho X), 
\end{equation}
where $X > 0$, we have 
\begin{eqnarray}
\widehat{K}_{(2)n}({\widehat {\bf \Pi}}, k)
=
{2\pi T D_{n}}
\int_{\rho_{\rm cut}}^{\infty} {d}\rho		
\left( 2 \sum_{\omega_{l}>0} e^{-2\omega_{l}\rho} \right)
\left< e^{2{i} \sigma I_{n} \rho} \right>_{\sigma}
\left< e^{2{i} J \rho \, {\rm sin}({k}/{2}) {\rm sin}k'} \right>_{k'}
\left< e^{{i} v_{n} \rho \, {\widehat {\bf p}}\cdot{\widehat{\bf \Pi}}} \right>_{\widehat {\bf p}}. 
\label{k2r}
\end{eqnarray}
Here, a lower cutoff $\rho_{\rm cut}$, corresponding to a high energy cutoff, of the $\rho$-integral has been introduced. 
Then, to average $\exp (i v_n \rho \, {\widehat {\bf p}}\cdot{\widehat{\bf \Pi}})$ with respect to the solid angle (or ${\widehat {\bf p}}$) in Eq. (\ref{k2r}), we introduce raising and lowering operators of the Landau level as $\widehat{\pi}_\pm
=\lambda_H(\widehat{\Pi}_x\pm{i}\widehat{\Pi}_y)/\sqrt{2}$. 
By expanding the exponential, we obtain 
\begin{eqnarray}
\left< e^{{i} v_{n} \rho \, {\widehat {\bf p}} \cdot \widehat{\bf \Pi}} \right>_{\widehat {\bf p}}
=
\exp\left(-\frac{{v_n}^2\rho^2}{4{\lambda_H}^2}\right)
\sum_{N=0}^{\infty}
\frac{1}{(N!)^2}\left(-\frac{{v_n}^2\rho^2}{2{\lambda_H}^2}\right)^N
\widehat{\pi}_+^N \widehat{\pi}_-^N. 
\label{eq:angleav}
\end{eqnarray}
From this equation, we see that the eigenstate of $\widehat{K}_{(2)n}({\widehat{\bf \Pi}}, k)$ is in the $n$-th Landau level.
Lastly, with respect to $\omega_l$, $\sigma$, and $k'$, we sum or average the functions expressed in Eq. (\ref{k2r}) and then, with Eq. (\ref{eq:angleav}), we find
\begin{eqnarray}
\widehat{K}_{(2)n}(\widehat{\bf \Pi}, k)
{\cal F}_N({\bf r})
&=& 
{2 \pi TD_{n}}
\int_{\rho_{\rm cut}}^{\infty} {d}\rho \,
\frac{{\rm cos} (2I_n\rho)}{{\rm sinh}(2 \pi T \rho)}
{\rm J}_{0}\left( 2J \rho \, {\rm sin}\frac{k}{2} \right) \nonumber \\
&\times& 
\exp\left(-\frac{{v_n}^2 \rho^2}{4{\lambda_H}^2}\right)
{\rm L}_N \left(\frac{{v_n}^2\rho^2}{2{\lambda_H}^2} \right)
{\cal F}_N({\bf r}), 
\label{eq:K2op}
\end{eqnarray}
where ${\cal F}_N({\bf r})$ belongs to the $N$-th Landau level, ${\rm J}_{0}(x)$ represents the zeroth-order Bessel function, and ${\rm L}_N(x)$ represents the $N$-th-order Laguerre polynomial.
Now by substituting Eq. (\ref{eq:K2op}) in Eq. (\ref{eq:Omega2}), we get $\Omega_{(2)}$ in the form of Eq. (\ref{secondOmega2}).

\subsection{Calculation of $\Omega_{(4)}$}

In this case, after performing the $\xi_n({\bf p})$ integral, the expression of $\widehat{K}_{(4)n}(\{{\widehat {\bf \Pi}}_i, k_i\})$ [Eq. (\ref{eq:K4})] consists of three denominators, and thus, the expression corresponding to Eq. (\ref{k2r}) of $\widehat{K}_{(4)n}(\{{\widehat {\bf \Pi}}_i, k_i\})$ is expressed as threefold integrals over the parameters $\rho_j$ ($j=1$, $2$, and $3$). After properly symmetrizing the expression w.r.t. $\widehat{\bf \Pi}_j$ and $k_j$, $\widehat{K}_{(4)n}(\{{\widehat {\bf \Pi}}_i, k_i\})$ takes the form 
\begin{eqnarray}
\widehat{K}_{(4)n}(\{\widehat{{\bf \Pi}}_i, k_i\}) 
&=& 
\frac{4 \pi T D_n}{N_{\rm layer}}
\int_0^\infty
{d}\rho_1 {d}\rho_2 {d}\rho_3 
\frac{{\rm cos}[2 I_n(\rho_1+\rho_2+\rho_3)]}{{\rm sinh}[2 \pi T(\rho_1+\rho_2+\rho_3)]} 
{\rm J}_0 \biggl(2 J \sqrt{D} \biggr) 
\nonumber \\
&\times& \langle \, \exp{[ 
{i} v_n {\widehat {\bf p}}\cdot(\rho_1{\widehat{\bf \Pi}}_1^* + \rho_2{\widehat{\bf \Pi}}_2 + \rho_3{\widehat{\bf \Pi}}_3^*) ]
} \, 
\rangle_{\widehat{\bf p}}, 
\label{eq:K4d}
\end{eqnarray}
where
\begin{eqnarray}
D(\{k_i, \rho_i\}) 
&=& 
\sum_{i=1}^3
\left[\rho_i \, {\rm sin}\biggl(\frac{k_i}{2}\biggr) \right]^2
+ \,\, 
2 \sum_{(i,j)=(1,2),(2,3)}
\,\, 
\rho_i\rho_j \, {\rm cos}\biggl(\frac{k_{ij}}{2}\biggr) {\rm sin}\biggl(\frac{k_i}{2}\biggr) {\rm sin}\biggl(\frac{k_j}{2}\biggr) \nonumber \\
&+& 
2\rho_1\rho_3 \, {\rm cos}\biggl(\frac{k_{24}}{2}\biggr) {\rm sin}\biggl(\frac{k_1}{2}\biggr) {\rm sin}\biggl(\frac{k_3}{2}\biggr) 
\end{eqnarray}
and $k_{ij}=k_i-k_j$.

Now, what we have to do is calculate the average of $\exp{[ 
{i} v_n {\widehat {\bf p}}\cdot(\rho_1{\widehat{\bf \Pi}}_1^* + \rho_2{\widehat{\bf \Pi}}_2 + \rho_3{\widehat{\bf \Pi}}_3^*) ]
}$ over the direction of ${\widehat {\bf p}}$ expressed in Eq. (\ref{eq:K4d}). 
Below, we adopt the Landau gauge (${\bf A}=Bx {\widehat y}$) and assume that, on the transition line, the order parameter belongs to the LLL and has only `$-$ component' (see Sect. 2). 
Then, we can expand the order parameter as $\Delta_n({\bf r}, k)= ({ S}_H)^{-1/2} w_{-,n}^{(0)}(k;T,H) \sum_q u_q({\bf r}) \phi(q,k)$, where 
$
{ S}_H 
= 
\sqrt{\pi} \lambda_H L_{y} 
$
with the system size $L_y$ along the $y$-axis, 
$
u_q({\bf r})
=
\exp[- (x+q{\lambda_{H}}^2)^2/(2 \lambda_H^2) + {i} q y]$, 
and $\phi(q,k)$ is the expansion coefficient.
We use the following identity:
\begin{equation}
\exp
\Big(
{i}\rho_i v_n {\widehat {\bf p}}\cdot\widehat{\bf \Pi}_i
\Big) 
u_{q_i}({\bf r}_i)
= \exp\biggl[- 
\frac{
(|\beta_{n,i}|^2-{\beta_{n,i}}^2)}{4} - \frac{
(x_i/\lambda_H+q_i \lambda_H + \beta_{n,i})^2}{2} + {i} q_i y_i
\biggr], 
\label{eq:identity}
\end{equation}
%%% We correct beta_n,i in 2/18. 
where $\beta_{n,i}=\rho_i v_n(\widehat{p}_x-{i}\widehat{p}_y)/\lambda_H$. 
We can show this by using the Campbell-Baker-Hausdorff formula.

Finally, by substituting Eq. (\ref{eq:K4d}) in Eq. (\ref{4th}), expanding the order parameter $\Delta_n({\bf r}_i,k_i)$ with $u_{q_i}({\bf r}_i)$, and using Eq. (\ref{eq:identity}), we can explicitly conduct the integration in Eq. (\ref{4th}) with respect to $x$ and $y$, and then obtain the final form of the quartic term 
\begin{eqnarray}
\Omega_{(4)}\Big|_{\rm LLL}
&=&
\frac{1}{\sqrt{2}{N_{\rm layer}} S_H}
\sum_{n,\{q_i\},\{k_i\}}
\left(\prod_{i=1}^4w_{-,n}^{(0)}(k_i;T,H)\right)
\delta_{k_1+k_3,k_2+k_4}
\delta_{q_1+q_3,q_2+q_4}
e^{-({q_{13}}^2+{q_{24}}^2){\lambda_H}^2/4}
\nonumber \\
&\times& 
\phi^*(q_1,k_1)
\phi(q_2,k_2)
\phi^*(q_3,k_3)
\phi(q_4,k_4)
\, 
2 \pi T D_n
\int_0^\infty
{d}\rho_1 {d}\rho_2 {d}\rho_3 
\frac{{\rm cos}[2I_n(\rho_1+\rho_2+\rho_3)]}{{\rm sinh}[2\pi T(\rho_1+\rho_2+\rho_3)]} \nonumber \\
&\times& {\rm J}_0\left(2J\sqrt{D}\right)
\biggl\langle 
e^{B}\Big|_{\rho_4=0} 
\biggr\rangle_{\widehat {\bf p}},
\label{eq:Omega4d}
\end{eqnarray}
where $q_{ij}=q_i-q_j$,
\begin{eqnarray}
B(\{q_i, \beta_{n,i}\})
&=& 
-\frac{1}{4}\left[\sum_{i=1}^4|\beta_{n,i}|^2-({\beta_{n,1}^*}^2+{\beta_{n,2}}^2+{\beta_{n,3}^*}^2+{\beta_{n,4}}^2)\right] \nonumber \\
&+& \frac{1}{8}
\left[
\left( 
\sum_{i=1}^4 \zeta_{n,i} \right)^2 - 4 \sum_{i=1}^4{\zeta_{n,i}}^2 
\right] 
+ \frac{1}{4}({q_{13}}^2+{q_{24}}^2){\lambda_{H}}^2,
\label{eq:B}
\end{eqnarray}
$\zeta_{n,i}=q_i \lambda_H - {\beta_{n,i}}^{*}$ $(i=1,3)$, and $\zeta_{n,i}=q_i\lambda_H + \beta_{n,i}$ $(i=2,4)$. 

Now, if we restrict the order parameter with modulation $\pm k$ along the $c$-axis, we obtain Eq. (\ref{nonlocal}) from Eq. (\ref{eq:Omega4d}).


\newpage
\begin{thebibliography}{9}
\bibitem{Bianchi1} A. Bianchi , R. Movshovich, C. Capan, A. Lacerda, P. G. Pagliuso, and J. L. Sarrao,  Phys. Rev. Lett. {\bf 91}, 
187004 (2003). 
\bibitem{Haga} F. Honda, R. Settai, D. Aoki, Y. Haga, T. D. Matsuda, N. Tateiwa, S. Ikeda, Y. Homma, H. Sakai, Y. Shiokawa, E. Yamamoto, A. Nakamura, and Y. Onuki, J. Phys. Soc. Jpn. Suppl.A {\bf 77}, 339 (2008). 
\bibitem{Tokiwa} Y. Tokiwa, R. Movshovich, F. Ronning, E. D. Bauer, A. D. Bianchi, Z. Fisk, and J. D. Thompson, Phys. Rev. B {\bf 82}, 220502(R) (2010). 
\bibitem{Kenzel14} S. Gerber, M. Bartkowiak, J. L. Gavilano, E. Ressouche, N. Egetenmeyer, C. Niedermayer, A. D. Bianchi, R. Movshovich, E. D. Bauer, J. D. Thompson, and M. Kenzelmann, Nat. Phys. {\bf 10}, 126 (2014). 
\bibitem{FF} P. Fulde and A. Ferrell, Phys. Rev.{\bf 135}, A550 (1964). 
\bibitem{LO} A. I. Larkin and Yu. N. Ovchinnikov, Sov. Phys. JETP {\bf 20}, 762 (1965). 
\bibitem{RI2} R. Ikeda, Phys. Rev. B {\bf 81}, 060510(R) (2008). 
\bibitem{HI15} Y. Hatakeyama and R. Ikeda, Phys. Rev. B {\bf 91}, 094504 (2015). 
\bibitem{Klein} U. Klein, Phys. Rev. B {\bf 69}, 134518 (2004); K. Yang and A. H. MacDonald, Phys. Rev. B {\bf 70}, 094512 (2004). 
\bibitem{AI03} H. Adachi and R. Ikeda, Phys. Rev. B {\bf 68}, 184510 (2003). 
\bibitem{RI1} R. Ikeda, Phys. Rev. B {\bf 76}, 134504 (2007). 
\bibitem{Kumagai06} K. Kumagai, M. Saitoh, T. Oyaizu, Y. Furukawa, S. Takashima, M. Nohara, H. Takagi, and Y. Matsuda, Phys. Rev. Lett. {\bf 97}, 227002 (2006). 
\bibitem{organics} S. Uji, T. Terashima, M. Nishimura, Y. Takahide, T. Konoike, K. Enomoto, H. Cui, H. Kobayashi, A. Kobayashi, H. Tanaka, M. Tokumoto, E. S. Choi, T. Tokumoto, D. Graf, and J. S. Brooks, Phys. Rev. Lett. {\bf 97}, 157001 (2006); H. Mayare, S. Kramer, M. Horvatic, C. Berthier, K. Miyagawa, K. Kanoda, and V. F. Mitrovic, Nat. Phys. {\bf 10}, 928 (2014). 
\bibitem{KFeAs} P. Burger, F. Hardy, D. Aoki, A. E. B{\" o}hmer, R. Eder, R. Heid, T. Wolf, P. Schweiss, R. Fromknecht, M. J. Jackson, C. Paulsen, and C. Meingast, Phys. Rev. B {\bf 88}, 014517 (2013). 
\bibitem{LiFeAs} K. Cho, H. Kim, M. A. Tanatar, Y. J. Song, Y. S. Kwon, W. A. Coniglio, C. C. Agosta, A. Gurevich, and R. Prozorov, Phys. Rev. B {\bf 83}, 060502 (2011). 
\bibitem{KFeSeEKFAs} V. A. Gasparov, A. Audouard, L. Drigo, A. I. Rodigin, C. T. Lin, W. P. Liu, M. Zhang, A. F. Wang, X. H. Chen, H. S. Jeevan, J. Maiwald, and P. Gegenwart, Phys. Rev. B {\bf 87}, 094508 (2013). 
\bibitem{Zocco} D. A. Zocco, K. Grube, F. Eilers, T. Wolf, and H. v. L{\"o}hneysen, Phys. Rev. Lett. {\bf 111}, 057007 (2013). 
\bibitem{Kasahara} S. Kasahara, T. Watashige, T. Hanaguri, Y. Kohsaka, T. Yamashita, Y. Shimoyama, Y. Mizukami, R. Endo, H. Ikeda, K. Aoyama, T. Terashima, S. Uji, T. Wolf, H. v. L{\" o}hneysen, T. Shibauchi, and Y. Matsuda, 
Proc. Natl. Acad. Sci. U.S.A. {\bf 111}, 16309 (2014). 
\bibitem{RIPRLCom} R. Ikeda, Phys. Rev. Lett. {\bf 102}, 069703 (2009).  
\bibitem{Agter} D. F. Agterberg, Z. Zheng, and S. Mukherjee, Phys. Rev. Lett. {\bf 100}, 017001 (2008). 
\bibitem{com1} Of course, depending on the value of $\Lambda_2$, there should be the situation in which the superconductivity near $T_{c0}$ is determined by band 1, while band 2 determines the SC properties in high fields. 
\bibitem{Gurevich} A. Gurevich, Phys. Rev. B {\bf 82}, 184504 (2010). 
\bibitem{Mizushima} T. Mizushima, M. Takahashi, and K. Machida, J. Phys. Soc. Jpn. {\bf 83}, 023703 (2014). 
\end{thebibliography}
\end{document}